# Studying a Surgery Service Occupation through a Queues Model

**Manuel Alberto M. Ferreira**

manuel.ferreira@iscte-iul.pt

Iscte-Instituto Universitário de Lisboa (ISCTE – IUL), ISTAR-IUL, Lisboa, Portugal

**Abstract**

A method to study and evaluate the occupation of a Hospital Surgery Service, with some specificity in its activity, is outlined in this work. Its application is exemplified with real data, and it is shown that it is simple, practical, and useful and allows a practical management of the service occupation.

## 1 Introduction

It is proposed in this work a method, based on a queues model, to study and evaluate the occupation of a Hospital (*H*) Surgery Service (*SS*), with characteristics quite frequent in this kind of services. The model here presented and developed is applied to real data about the activity of a *H SS* service in the first three Quarters in a determined year[1].

In section 2 the problem, joint with the *SS* characteristics, is described and Institutional data are presented. Then, in section 3, the queues model is described and the method - the main objective of this work - is outlined. In section 4 the model parameters are estimated; and in section 5 a discussion of the results is performed. The work ends with a Conclusions section.

## 2 The Problem

A *SS* at *H* operates patients admitted directly from outside or transferred from the other *H* services.

Surgical patients stay in the service in the postoperative period and then are discharged. There are no transferences from the *SS* to the other services of *H*. In fact, the operated patients were either followed in other services, and consequently transferred to the *SS*, or followed in the *SS* itself.

---

[1]All this secrecy on data is an Institutional condition to allow their use.

The *SS* politics is to hospitalize patients only when they have a scheduled surgery. Patients who will be operated in the *SS*, which surgery is not scheduled yet, wait for schedule either interned in the other *H* services or at home.

So, patients' arrivals at the service are conditioned by the operating room daily working time.

In *SS* there are 50 available beds for hospital patients. Here are not considered the available beds for *SS* in Intensive Care Unit (*ICU*) because when a patient is in *ICU* his/her bed in *SS* is considered as occupied by him/her.

Some *SS* staff elements are convinced that the service is under occupied. Such a convincing is based in the fact that there is, in general, a lot of unoccupied beds in *SS*. And it is partially corroborated by the *H* inner statistic numbers that give occupation indexes for the $1^{st}$, $2^{nd}$, and $3^{rd}$ Quarters of the year under study of 76.08%, 68.66% and 77.46%, respectively, see Table 2.4. Note that these percentages were calculated as the ratio of the daily mean number occupied beds over the total number of 50 beds by quarter.

Table 2-**Movement on inpatients at *SS* in the first three quarters of the year under study (available in *H* inner statistic numbers)**

Table 2.1-**Patients**

| Patients | $1^{st}$ Quarter | $2^{nd}$ Quarter | $3^{rd}$ Quarter |
|---|---|---|---|
| **Initial number** | 29 | 29 | 33 |
| **Coming from outside *H*** | **257** | **228** | **263** |
| **Transferred from other *H* services** | **168** | **138** | **166** |
| **Total cared** | 454 | 395 | 462 |
| **Total discharged** | 425 | 362 | 425 |

Table 2.2-**Deaths**

| Deaths | $1^{st}$ Quarter | $2^{nd}$ Quarter | $3^{rd}$ Quarter |
|---|---|---|---|
| **Total** | 6 | 11 | 9 |
| **By cared patient (%)** | 1.32 | 2.78 | 1.94 |
| **By discharged patient (%)** | 1.41 | 3.04 | 2.11 |

Table 2.3-**Hospital days**

| Hospital days | $1^{st}$ Quarter | $2^{nd}$ Quarter | $3^{rd}$ Quarter |
|---|---|---|---|
| **Total** | 3323 | 2999 | 3421 |
| **By cared patient (%)** | 7.35 | 7.59 | 7.40 |
| **Global mean delay** | 7.82 | 8.28 | 8.04 |

Table 2.4-**Occupation indexes**

| Occupation indexes | $1^{st}$ Quarter | $2^{nd}$ Quarter | $3^{rd}$ Quarter |
|---|---|---|---|
| **Cared patients by bed** | 8.85 | 7.54 | 8.85 |
| **Mean inoccupation by bed** | 2.46 | 3.78 | 2.34 |
| **Services occupation (%)** | 76.08 | 68.66 | 77.46 |

So, there having patients waiting for operation scheduling, those *SS* staff members ask if:

- In fact, the service will not be working with less than 50 beds. And, if so, with its hospital days capacity under used?
- Is it possible to increase the operating room daily time working, so increasing the inpatients arrivals at the service to decrease the number of patients waiting for schedule and increasing the use of the whole *SS* capacity occupation?

For the moment, with the practiced schedule rhythm, there is never a lack of beds: there are no losses in the sense that there is no need of not scheduling an operation due to the lack of a bed to install the patient. That is: the observed probability of the total 50 beds occupation is insignificant. But, the surgeries rhythm increase, evidently, would increase that probability that must be kept in controllable levels. This means, such that those losses occur rarely, and may be overcome with rare *SS* internship capacity increases, using stretchers for instance.

## 3 The Model

In [1] it is suggested that it is possible to evaluate an existing service with a finite capacity in a hospital impact, modelling it as a network of queues, each one with finite servers (beds) except the one corresponding to the service with a finite capacity. This will be modelled as a Centre with finite servers with losses. The customers (patients) that arrive at this service find the whole beds occupied, are lost to the system and directed to another service.

So, consider a hospital with $n$ services $1, 2, \ldots, n$. Call $p_{ij}$ the probability that a patient is transferred from service $i$ to service $j$, $i, j = 1, 2, \ldots, n$ and $\lambda_i$ the patients external arrivals rate at service $i$, $i = 1, 2, \ldots, n$. The various external arrivals processes are supposed to be Poisson processes.

Being $P = \begin{bmatrix} p_{11} & \cdots & p_{1n} \\ \vdots & \ddots & \vdots \\ p_{n1} & \cdots & p_{nn} \end{bmatrix}$ the commutation matrix, see [2,3,4 and 7], that describes the transitions among the various services, and $[\gamma_1\ \gamma_2\ \ldots\ \gamma_n]$ the total arrivals, internal and external, rates vector,

$$[\gamma_1\ \gamma_2\ \ldots\ \gamma_n] = [\lambda_1\ \lambda_2\ \ldots\ \lambda_n] + [\gamma_1\ \gamma_2\ \ldots\ \gamma_n] \begin{bmatrix} p_{11} & \cdots & p_{1n} \\ \vdots & \ddots & \vdots \\ p_{n1} & \cdots & p_{nn} \end{bmatrix} \quad (3.1).$$

Suppose that the service 1 is the service with a finite capacity. That is: suppose that this service has $c$ servers: $c$ beds. In [1] it is shown that in situations of

- Service times, that is: hospital days, exponentially distributed, see [5], or
- Service times with any distribution but without feedback, see [2 and 5][2],

---

[2] The feedback lack means the impossibility of a customer to come back to unity 1 after leaving it.

the probability that the service 1 is fully occupied is given by the Erlang loss formula, see for instance [1 and 3]:

$$\alpha = \frac{\frac{\rho^c}{c!}}{\sum_{j=0}^{c} \frac{\rho^j}{j!}} \qquad (3.2)$$

where

$$\rho = \frac{\gamma_1}{\mu_1} \qquad (3.3)$$

is the traffic intensity, being $\gamma_1$ given by equation (3.1) and $\mu_1^{-1}$ the service mean hospital days.

In the case of service times with any distribution with feedback, it is suggested in [1], because of computational simulations, that formula (3.2), although not exact, is a good approximation for the service 1 fully occupation probability. The authors emphasize that formula (3.2) is not exact due to the service 1 does not behave as a $M|G|c$ queue with losses since, in these circumstances, there is no guarantee that the total arrivals process is a Poisson process.

The problem under study in this work will not be approached in this way, because the patients that ask for hospital days in the $SS$ flow is determined by the operating room daily activity time instead of the external arrival's flows.

Consequently, formula (3.2) will be used as follows[3]:

- It is determined a value for $\rho$, called $\rho_s$ ($\rho$ standard) after a situation where losses are not observed (as it is the case at $SS$). It is possible to use for it the Little's formula, see for instance [3], in the form:
$$E[n] = \rho \qquad (3.4)$$
because it is assumed that there are no losses and it may be considered that $SS$ behaves as an infinite server's queue,

- Using $\rho_s$ and $c = 50$ in (3.2) it is determined a value $\alpha_s$ ($\alpha$ standard) that is taken as the value of $\alpha$ to which in practice correspond no losses,
- After $\alpha_s$ it is refereed a maximum value admissible for $\alpha$. In the present study it will be considered $10\alpha_s$,
- It will be evaluated the really used hospital days capacity calculating the value $i$ such that $\alpha^i < 10\alpha_s$, being:

---
[3] This is the method, based on a queues model, to study and evaluate the occupation of a Hospital (H) Surgery Service (SS), with particular characteristics.

$$\alpha^i = \frac{\frac{\rho_s^{50-i}}{(50-i)!}}{\sum_{j=0}^{50-i} \frac{\rho^j}{j!}}, i = 0,1,2,\dots \quad (3.5).$$

Note that $\alpha^0 = \alpha_s$,

- It will be evaluated the percentage in which it is possible to increase the number of scheduled operations, calculating the value $k$ such that $\alpha_k < 10\alpha_s$, where:

$$\alpha_k = \frac{\frac{(k\rho_s)^{50}}{50!}}{\sum_{j=0}^{50} \frac{(k\rho_s)^j}{j!}}, k = 1; 1.05; 1.10; \dots \quad (3.6).$$

Note that $\alpha_1 = \alpha_s$.

Note by its turn that, in accordance with [1], these formulae are exact if the number of scheduled operations by day in *SS* is a Poisson process. Even if this does not happen, they are good approximations. Anyway, in this concrete situation, there are not enough data available to test the Poisson process hypothesis.

## 4 The Parameters Estimation

The estimation of $E[n]$ was performed calculating the ratio of the total Hospital days, see Table 2.3, in the first three Quarters of the year over 274, the first 9 months of the year number of days. It was obtained $\rho_s = 35.55839416^4$. In Table 4.1 the values of $\alpha^i/\alpha^0, i = 1,2,\dots,10$ are presented:

Table 4.1 - $\alpha^i/\alpha^0, i = 1, 2, \dots, 10$

| $i$ | $\alpha^i$ | $\alpha^i/\alpha^0$ |
|---|---|---|
| 0 | 0.43% | 1 |
| 1 | 0.60% | 1.4 |
| 2 | 0.83% | 1.93 |
| 3 | 1.13% | 2.63 |
| 4 | 1.51% | 3.51 |
| 5 | 1.99% | 4.63 |
| 6 | 2.56% | 5.95 |
| 7 | 3.26% | 7.58 |
| 8 | 3.57% | **8.30**(<10) |

---

[4] So directly observed values were used and the values calculated after the observed ones were avoided.

| | | |
|---|---|---|
| 9 | 5.01% | 11.65 |
| 10 | 6.01% | 14.19 |

In Table 4.2 the values of $\alpha_k/\alpha_1$, $k = 1; 1.05; ...; 1.40$ are presented:

Table 4.2- $\alpha_k/\alpha_1$, $k = 1; 1.05; ...; 1.40$

| k | $\alpha_k$ | $\alpha_k/\alpha_1$ |
|---|---|---|
| 1 | 0.43% | 1 |
| 1.05 | 0.83% | 1.93 |
| 1.10 | 1.47% | 3.42 |
| 1.15 | 2.37% | 5.51 |
| **1.20** | 3.54% | **8.23**(<10) |
| 1.25 | 4.98% | 11.58 |
| 1.30 | 6.65% | 15.47 |
| 1.35 | 8.51% | 19.79 |
| 1.40 | 10.51% | 24.44 |

## 5 Discussion

It is clear, according to the chosen criterion, that:

- From Table 4.1, with 42 beds (note the values in bold at Table 4.1), in the present conditions, the *SS* performance should not be substantially different from the actual.

So, the *SS* real internship capacity should be estimated in 84%, greater than the 74% that is the *SS* weighted mean for the three first Quarters,

- From Table 4.2, an increase of 20% (note the values in bold at Table 4.2) in the operating room working time[5] should be accommodated with no problems by *SS*.

This supports the staff opinion of *SS* under occupation. Really, with 42 beds the calculated *SS* full occupation probability is 3.57%. So, there having had 1220[6] demands for hospital days in the first three Quarters of the year, this means that it should be expected the existence of problems in 44 cases (a monthly mean of 5).

Increasing the scheduling rhythm in 20%, there should have 1464 demands for hospital days in the same period. And the expected existence of problems should be in 52 cases (a monthly mean of 6).

So, eventually, 5 or 6 times per month an extra bed, a stretcher, should be necessary, in cases of reduction of the number of beds or a scheduling increase, respectively, considered.

---

[5] Admitting that the number of hospital days is directly proportional to the operating room working time.
[6] See numbers in bold at Table 2.1.

Of course, the use of this model requires small values of $\alpha_s$. In the situation here studied it was 0.43%. If this does not happen the hypothesis of an observed situation with no losses must be discarded.

## 6 Conclusions

The method presented in this study is obviously simple, practical, and useful. It relies on some credible assumptions about the queues used to model the Hospital services behaviour and on the admissible ranges for $\alpha^i$ and $\alpha_k$, arbitrary as it was pointed in the text, being determined after the manager experience and the risk judged admissible.

## Acknowledgments

This work is financed by national funds through FCT - Fundação para a Ciência e Tecnologia, I.P., under the project FCT UIDB/04466/2020. Furthermore, the author thanks the Iscte-Instituto Universitário de Lisboa and ISTAR-IUL, for their support.

## List of Acronyms

*H*-Hospital

*SS*-Surgery Service

*ICU*-Intensive Care Unit